\documentstyle[12pt]{article}
\begin{document}
\vskip 4 cm
\begin{center}
\Large{\bf UNIVERSAL FORCES AND THE DARK ENERGY PROBLEM}
\end{center}
\vskip 3 cm
\begin{center}
{\bf AFSAR ABBAS} \\
Centre for Theoretical Physics\\
JMI, Jamia Nagar, New Delhi - 110025, India\\
(e-mail : safsarabbas@gmail.com)
\end{center}
\vskip 20 mm  
\begin{centerline}
{\bf Abstract }
\end{centerline}
\vskip 3 mm

The Dark Energy problem is forcing us to re-examine our 
models and our understanding of relativity and space-time. 
Here a novel idea of Fundamental Forces is introduced. 
This allows us to perceive the General Theory of Relativity 
and Einstein's Equation from a new pesrpective. 
In addition to providing us with an improved understanding of 
space and time, it will be shown how it leads to a 
resolution of the Dark Energy problem.

\newpage

Dark Energy is certainly the most puzzling problem in physics and 
astronomy today [1]. All kind of proposals to solve the problem 
are being put forward, but we are nowhere near a resolution of 
the issues involved. In this paper the author suggests 
incorporation of a new concept in our understanding of Nature 
and which helps in explaining the Dark Energy problem 
and has the potential of providing us with an improved 
understanding of Nature.

The new concept is called "Universal Force". It was first proposed 
by philosopher-scientist Hans Reichenbach [2]. Although this novel 
concept was actually suggested in what may be termed as the  
philosophical context, here the author would like to emphasize as 
to how the same can be used as a powerful tool in the 
physical context as well. Actually we shall see that we have to 
improve upon the original idea of Reichenbach [3] 
in a significant manner to be able to use 
it in physics and astronomy. As to what changes are necessary 
would be discussed below after the original idea of the Universal 
Forces of Reichenbach is introduced.

To the sceptics who may believe that philosophy of science has no 
relevance to the actual science itself, it may be pointed 
out that one 
never loses by being open to ideas from any quarters whatsoever.
Obviously the only criteria would be relevance and proper 
applicability of the idea in whatever science one is talking 
about. A recent Conference Proceeding [4] would attest to this 
fact. In the case of the Dark Energy conundrum, wherein 
we do not even know where we stand, 
this point becomes all the more pertinent.

Within the discipline of philosophy of science the Universal 
Force concept of Reichenbach has had mixed reactions. While a few 
philosophers have been supportive of the concept [5,6,7],
some others have been critical of it [8,9,10]. We shall not delve 
too deep into the philosophical issues of the same, as that 
would take us farther away from our main purpose here and 
which is to see how to properly utilize the concept as 
scientists. Hence we shall use only those points as would be found 
to be relevant for our purpose here. It would suffice to mention 
here as to what Rudolf Carnap has stated in the Introduction of 
Reichenbach's book [3]. He called the concept as " ... of great 
interest for the methodology of physics but what has so far 
not received the attention it deserves".
In this paper we shall try to rectify for this failure of 
appreciating the concept of the Universal Force - albeit
in a somewhat altered and improved form.

Reichenbach defines two kind of forces - Differential Forces and 
Universal Forces. It may be pointed out that 
the term "force" here should not be taken strictly as
defined in physics but in a broad and general framework. 
In fact Carnap has suggested that the term "effect" 
instead of "force' would better serve the purpose [5]
and which allows it be used in different frameworks. 
Hence to 
conform with the accepted practice, though in this paper we 
shall continue to use the term "Universal Force" the reader may do 
well to remember that what we really mean is "Universal Effect".

One calls a force Differetial if it acts differently on different 
substances. It is called Universal if it is quantitatively the 
same for all the substances [3,5]. If we heat a rod of initial 
length $l_0$ from initial temperature $T_0$ to tempetature T then 
its length is given as

\begin{equation}
l = l_0 [ 1 + \beta ( T - T_0 ) ]
\end{equation}

where $\beta$ the coefficient for thermal expansion is different 
for different materials. Hence this is a Differential Force.
Now the correction factor due to the influence of gravitation on 
the length of the rod is

\begin{equation}
l = l_0 [ 1 - C { m \over r } {cos{^2} \phi}]
\end{equation}

Here the rod is placed at a distance r from sun whose mass is m 
and $\phi$ is the angle of the rod with respect to the the line 
sun to rod. C is a universal constant ( in CGS unit 
C= 3.7 x ${10}^{-29}$ ). As this acts in the same manner for any 
material of mass m, gravity is a Universal Force
as per the above definition.

Reichenbach also gives a general definition of the Universal 
Forces [3,p 12] as: (1) affecting all the materials in the same 
manner and (2) there are no insulating walls against it. We saw 
above that gravity is such a force,

Indeed gravity is a Universal Force par excellance. It affects all 
matter in the same manner. The equality of the gravitational and 
inertial masses is what ensures this physically. If the 
gravitational and inertial masses were not found to be equal, then 
one would not have been able to visualize of the paths of freely 
falling mass points as geodesics in the four dimentional space-time.
In that case different geodesics would have resulted from 
different materials of mass points [3]. 

Therefore the universal effect of gravitation on different kinds 
of measuring instruments is to define a single geometry for all of 
them. Viewed this way, one may say that gravity is 
geometerized. "It is not theory of gravitation that becomes 
geometry, but it is geometry that becomes the experience of the 
gravitational field" [3, p 256]. Why does the planet follow the 
curved path? Not because it is acted upon by a force but 
because the curved space-time manifold leaves it with no other 
choice!

So as per Einstein's theory of relativity, one does not speak of 
a change produced by the gravitational field in the measuring 
instruments, but regard the measuring instruments as free from any 
deforming forces. Gravity being a Universal Force, in the 
Einstein's Theory of Relativity, it basically disappears and is 
replaced by geometry.

In fact Reichenbach [3, p 22] shows how one can give a consistent 
definition of a rigid rod - the same rigid rods which are needed 
in relativity to measure all lengths. "Rigid rods are solid bodies 
which are not affected by Differential Forces, or concerning which 
the influence of Differential Forces has been eliminated by 
corrections; Universal  Forces are disregarded. We do not neglect 
Universal Forces. We set them to zero by definition. Without such 
a rule a rigid body cannot be defined." In fact this rule also 
helps in defining a closed system as well. 

All this was formalized in terms of a theorem by 
Reichenbach [3, p 33]

\vskip 1 cm

{\bf THEOREM $\theta$} :

Given the geometry $G^0$ 
to which the measuring instruments 
conform, we can imagine a Universal Force F which affects 
the instruments in such a way that the actual geometry is an 
arbitrary geometry $G$, while the observed deviation from $G$
is due to universal deformation of the measuring instruments."

\begin{equation}
{G^0} + F = G 
\end{equation}

Hence only the combination ${G^0} + F$ is testable. 
As per Reichenbach's 
principle one prefers the theory wherein we put F=0.
If we accept Reichenbach principle of putting the 
Universal Force of gravity to zero, then the arbitrariness in the 
choice of the
measuring procedure is avoided and the question of the geometrical 
structure of the physical space has a unique answer determined by 
physical measurement. It is this principle which Carnap praises 
highly [5, p 171], " Whenever there is a system of physics in 
which a certain universal effect is asserted by a law that 
specifies 
under what conditions in what amount the effect occurs, then the 
theory should be transformed so that the amount of effect would be 
reduced to zero. This is what Einstein did in regard to 
contraction and expansion of bodies in gravitational field."
The left hand side of Einstein's equation (below)
gives the relevant non-Euclideon geometry

\begin{equation}
G_{\mu \nu} = 8 \pi G \langle \phi | T _{\mu \nu} | \phi \rangle
\end{equation}

In the case of gravity, and in as much as Einsteins's Theory of 
Relativity has been well tested experimentally, we treat the 
above concept as well placed empirically. But from this single 
success Reichenbach generalizes this as a fundamental principle 
for all cases where Universal forces may arise. As Carnap states 
[5, p 171], " Whenever universal effects are found in physics, 
Reichenbach maintained that it is always possible to eliminate 
them by suitable transformation of theory; such  
a transformation should be made because of the overall simplicity 
that would result. This is a useful general principle, 
deserving more attention than it has received.
It applies not only to relativity theory, but also to situations 
that may arise in the future in which other universal effects may 
be observed. Without the adoption of this rule there is no way to 
give unique answer to the question - 
what is the structure of space?". 

As such Reichenbach goes ahead and tries to apply this principle 
of elimination of Universal Forces to another universal effect 
that he finds and which arises from considerations of 
topology ( as an additional consideration over and above that of 
geometry ) of space-time of the universe.

The Theorem $\theta$ is limited to talking about the geometry of
space-time only. It does not take account of specific 
topological issues 
that may arise. To take account of topology of the space-time 
we shall have to extend the said theorem appropriately.

What would one experience if space had different topological 
properties. To make the point home Reichenbach considers a 
torus-space [3, p 63]. This is quite detailed and extensive.
However for the purpose of simplifying the
and shortening the discussion here we shall
talk of a two dimensional being who lives on the 
surface of a sphere. His measurements tell him so. But in
spite of this he insists that he lives on a plane. 
He may actually do so as per our discussion above if he confines 
himself to metrical relations only.
With an appropriate Universal Force he can he can justify living
on a plane. But the surface of a sphere is topologically 
different 
from that of a plane. On a sphere if he starts at a point X and 
goes on a world tour he may come back to the same point X. But 
this is impossible on a plane. And hence 
to account for coming back to the "same point" 
he has to maintain that on the plane he 
actually has come back to a different point Y - which though is 
identical to X in all other respects. 
One option for him is to 
accept that he is actually living on a sphere. 
However if he still wants 
to maintain his position that he is living on a plane then he has 
to explain as to how point Y is 
physically identical to point X in spite of 
the fact that X and Y are different and distinct points of space. 
Indeed he can do so by visualizing a fictitious force
as an 
effect of some kind of "pre-established harmony" [3, p 65] by 
proposing that everything that occurs at X also occurs at the 
point Y. As it would affect all matter in the same manner this 
corresponds to a Universal Force/Effect as per Reichenbach's 
definition.

This interdependence of corresponding points which is essential 
in this "pre-established" harmony cannot be interpreted as 
ordinary causality, as it does not require ordinary time to 
transmit it
and also does not spread continuously through intervening space. 
Hence there is no mysterious causal connection between the points 
X and point Y. Thus this necessarily entails 
proposing a "causal anomaly" [3, p 65].
In short connecting different topologies through a fictitious
Universal Effect of "pre-established harmony" necessarly calls 
for introduction of "causal anomalies". 
Call this new hypothesize 
Universal Force as A and the Theorem $\theta$ be extended to 
read

\begin{equation}
{G^0} + F  + A = {\bf G} 
\end{equation}

where on the right had side we have given a different 
capital ${\bf G}$ 
which reduces to $G$ of the original Theorem $\theta$
when A is set equal to zero.

Now as per Reichenbach's law of preferring that physical reality 
wherein all Universal Forces are put to zero, he 
advocates of putting A to zero. He pointed out that this has the 
advantage of retaining physical "causality " in our science, 
This he takes as  a success of his methodology. As per Reichenbach 
[3, p 65] " The principle of causality is one of its (physics)
sacred laws, which it will not abandon lightly; pre-established 
harmony, however is incompatible with this law".

However, as the said 'causal anomaly" is of topological origin we 
cannot be sure in what manner it will manifest itself physically. 
In addition
will not the Universal Force/Effect of "pre-established harmony"
compensate for it in some manner?
So what one is saying is that it is possible that Reichenbach was 
wrong in putting all Universal Forces to zero. It was OK to put F 
to zero which justified the geometrical interpretation of 
gravity.
But in the case of this new topological Universal Force we 
really do not know enough 
and let us not be governed by any theoretical prejudice and 
let the Nature decide as to what is happening. So to say, let us 
look at modern cosmology to see if it is throwing up any new 
Universal Forces which may be identified with our 
"pre-established harmony" here.

To understand this let us look at the Einstein's Equation given 
above. Harvey and Schucking [11] correcting for Einstein's error 
in understanding the role of the cosmological term $\lambda$
have derived the most general equation of motion to be

\begin{equation}
G_{\mu \nu} + \lambda {g_{\mu \nu}} 
= 8 \pi G \langle \phi | T _{\mu \nu} | \phi \rangle
\end{equation}
 
They showed that [11] the Cosmological Constant $\lambda$ 
above provides 
a new repulsive force proportional to mass m, repelling every 
particle of mass m with a force

\begin{equation}
F = m {c^2} {\lambda \over 3} x
\end{equation}

Recent data [1] on $\lambda$ is what leads to the crisis of 
Dark Energy.

Quite clearly this repulsive force
is a new Universal Force as per our definition
and hence conforms to the "pre-established harmony" aspect of the 
"causal anomaly".
Thus we see that indeed as per the recent data on accelerating 
universe we have stumbled upon this new Universal Force which is 
of topological origin. Hence the source of dark energy is 
due to "causal anomaly" arising from the unique topological 
structure of our universe.
This solves the mystery of the origin of Dark Energy.

So we would like to emphasize that it is the accelerating 
universe ( and hence the Dark Energy ) 
which is forcing us to accept the incorporation of this
"causal anomaly" of topological origin.  
Implications of this new concept in physics have now to be 
explored.

Note that as per Theorem $\theta$ when one puts F to be zero then 
one obtains the proper non-Euclidean Geometry of Einstein's 
equation.
But now we know that full structure is the sum of this 
non-Euclidean geometry plus A , the new
Universal Force ( as per the modified theorem above )
and this is what the accelerating universe is forcing us to 
accept.
This is what we called capital ${\bf G}$ above.
We feel that the DASI data on $\Omega_0$ being close to one 
and thus showing that the Universe is flat [1] is consistent with
capital ${\bf G}$ being equal to ${G + A}$. 
In principle just as per the original Theorem $\theta$
one may add a Universal Force F to Einstein's non-Euclidean 
geometry 
to obtain a physically relevant Euclidean geometry, 
so in the same manner given a non-Euclidean geometry of 
Einstein on can add an appropriate Universal Force A to provide
a flat universe. And this is exactly what capital ${\bf G}$ is 
telling us. Thus the observed flatness of the universe  
may be treated as a success of the new idea proposed here.

One would like to ask as to in what other manner incorporation of 
this new "causal anomaly" may help us in understanding Nature 
better? Will it provide new perspectives as 
answers to quantum mechanical puzzles of 
quantum jumps, non-locality etc. These are open questions  
to be tackled in future.

\newpage

\vskip 1 cm
\begin{center}
{\bf REFERENCES }
\end{center}

\vskip 1 cm

1. M S Turner, "Making sense of the new cosmology", 
{\it Int J Mod Phys}, {\bf A17S1} (2002) 180-196

2. Hans Reichenbach (1891-1953) can properly be called a 
philosopher-scientist. As a leading philosopher of science
he was founder of the Berlin Circle and a proponent of logical 
positivism. Among his teachers were David Hilbert, Max Planck, 
Max Born and Albert Einstein. He wrote extensively on the theory 
of probability, theory of relativity and quantum mechanics.
His philosophical writings have a definite scientific touch in 
them, very much akin to that of Descartes, Leibniz and Huygens.

3. H Reichenbach, "The philosophy of space and time", Dover, 
New York (1957) (Original German edition in 1928)

4. C Callender and N Huggett, " Physics meets philosophy at
the Planck scale", Cambridge University Press, UK (2001)

5. R Carnap, "An introduction to the philosophy of science",
Basic Books, New York (1966)

6. E Nagel, "The structure of science", Routledge and 
Kegan Paul, London (1961)

7. D Dieks, "Gravitation as a Universal Force", 
{\it Synthese}, {\bf 73} (1987) 381-397

8. B Ellis, "Universal and Differential Forces", 
{\it Brit J Phil Sc}, {\bf 14} (1963) 177-194

9. A Gruenbaum, "Philosophical problems of space and time", 
Dordrecht, Holland; D Reidel (1973) or
Alfred A Knopf, New York (1963)

10. R Torretti, "Relativity and geometry", Pergamon Press (1983)

11. A Harvey and E Schucking, "Einstein's mistake and the 
cosmological constant", 
{\it Am J Phys}, {\bf 68} (2000) 723-727

\end{document}